\documentstyle[prd,aps,epsf]{revtex}
\begin{document}

\draft \preprint{v7.2-Final}

\title{Navigation in Curved Space-Time}
\author{Thomas B. Bahder}
\address{U. S. Army Research Laboratory \\
2800 Powder Mill Road \\
Adelphi, Maryland, USA  20783-1197}

\date{\today}
\maketitle

\begin{abstract}
A covariant and invariant theory of navigation in curved
space-time with respect to electromagnetic beacons is written in
terms of J. L. Synge's two-point invariant world function.
Explicit equations are given for navigation in space-time in the
vicinity of the Earth in Schwarzschild coordinates and in rotating
coordinates.  The restricted problem of determining an observer's
coordinate time when their spatial position is known is also
considered.
\end{abstract}


\section{Introduction}
Curved space-time forms the basis for most classical theories of
gravity, such as general relativity.  These theories are usually
based on a metric for four dimensional space-time~\cite{Will}.
Some of the basic concepts used in general relativity and related
theories are transformation rules for tensors, the affine
connection, and the relation of the metric to proper time along an
observer's world line.  A useful, but little-used concept, is that
of the world function of space-time, as developed by J. L. Synge.
The world function is essentially one-half of the squared measure
between two points in space-time. The utility of the world
function comes from the fact that it is closely related to
experiments, and that it is a type of scalar quantity. Since the
world function transforms as a kind of scalar, it allows us to
formulate geometric quantities in a covariant way. Hence, the
world function is a valuable tool for understanding the geometric
ideas in metric theories of gravity, in three dimensional
differential geometry and tensor analysis, and wherever arbitrary
coordinate systems are used. As an example of the utility of the
world function, I present its application to the problem of
navigation in a curved space-time.  This application actually goes
beyond a simple pedagogical example because it deals with the
real-world need for precise navigation and time dissemination.

Consider the problem of an observer who wants to navigate in a
curved space-time with respect to electromagnetic beacons. I use
the word navigate to mean that the observer determines his or her
coordinate position and coordinate time along their world line, in
some system of space-time coordinates. I assume that the
electromagnetic beacons continuously broadcast their space-time
coordinates and that this information is imbedded in the emitted
electromagnetic signals. Furthermore, I assume that an observer at
unknown coordinates $ (t_o,{\bf x}_o ) $ simultaneously receives
these signals from four beacons.  The observer's navigation
problem is to compute his position $ (t_o,{\bf x}_o ) $ from the
four received emission event coordinates $ (t_s,{\bf x}_s ) $,
$s=$1,2,3,4, of the electromagnetic beacons.

\begin{figure}[htbp]
\centerline{\epsfsize=6.0cm \epsfbox{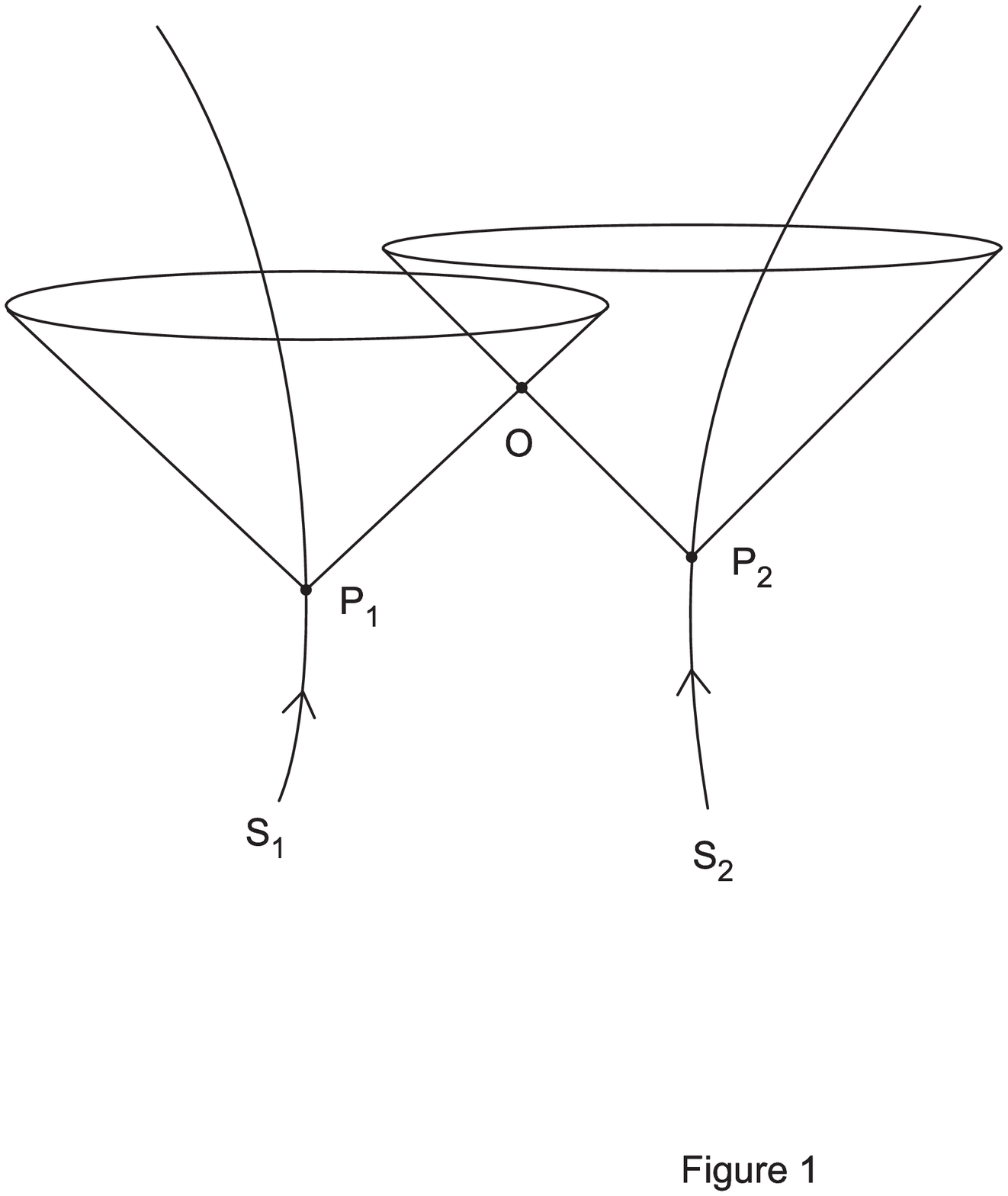}}
\caption{Space-time diagram showing two (of the four)
satellites with world lines $S_1$ and $S_2$. Electromagnetic
signals are emitted at $P_1$ and $P_2$,
and reach the observer at $O$.}
\label{LightConeIntersectionFigure}
\end{figure}

In the case of flat space-time, the observer must solve the four
simultaneous equations
\begin{equation}
 | {\bf x}_o - {\bf x}_s |^2 - c^2 (t_o - t_s )^2  =0,
 \; \; \; \;  s=1,...4
\label{LightCone}
\end{equation}
Equation (\ref{LightCone}) contains four invariant statements:
light signals travel `on the light cone' from each emission event
to the observer, (see Figure \ref{LightConeIntersectionFigure}).
To resolve the branches of the light cones, the causality
conditions $t_o > t_s$, for $s=$1,$\dots$4, must be added.  The
relations in Eq.\ (\ref{LightCone}) are the basic navigation
equations applied by users of satellite navigation  systems, such
as the U.S. Global Positioning System (GPS) and the Russion Global
Navigation Satellite System
(GLONASS)~\cite{ParkinsonGPSReview,Kaplan96,Hofmann-Wellenhof93}.
The coordinates of the four events, $(t_s, {\bf x}_s)$, correspond
to particular radio emissions by the satellites. The emission
event coordinates can be extracted from information transmitted by
digital codes.

Equation (\ref{LightCone}) is commonly used in two different ways.
First, an observer may receive radio signals from satellites and
compute his  or her space-time coordinates $(t_o, {\bf x}_o)$ in
terms of four known satellite emission events $(t_s, {\bf x}_s)$.
The second use of Eq. (\ref{LightCone}) is to locate a satellite,
which is at an unknown position $(t_o, {\bf x}_o)$ in terms of
four ground observations at known coordinates $(t_s, {\bf x}_s)$.
For this use, we apply the causality condition $t_s
> t_o$, $s=$1,$\dots$4.  In Eq. (\ref{LightCone}), the assumption is made
that space-time is flat and the speed of light $c$ is constant.
Furthermore, by using Eq.\ (\ref{LightCone}) we make the geometric
optics approximation that the wavelength of the electromagnetic
waves is small compared to all physical dimensions of the receiver
and transmitter systems~\cite{MTW,KravtsovOrlov}.

In recent years, there have been significant improvements in the stability of
frequency standards and measurement techniques~\cite{PetitWolf94,WolfPetit95}.
Consequently, over satellite-to-ground distances, precise measurements should be
interpreted within the framework of a curved space-time
theory~\cite{Guinot97,GAIA,Bahder98FermiCoord,Sovers}. Furthermore, the equations
for navigation in space-time should be manifestly covariant and also
invariant~\cite{Kheyfets}.

In this work, I write down a generalization of the navigation Eq.\
(\ref{LightCone}) for curved space-time and give the detailed
equations that must be solved for navigating in the vicinity of
the Earth, both in Schwarzschild coordinates and in rotating
coordinates. I still retain the geometric optics approximation,
however, I take into account deviations from flatness to first
order in the metric. This means that the flat-space light cones in
Eq.\ (\ref{LightCone}) are replaced by equations for null
geodesics. The required navigation equations are simply
expressible in terms of the world function developed by J. L.
Synge~\cite{Synge1960}. The resulting formalism takes into account
the delay of electromagnetic signals due to the presence of a
gravitational field. The detailed equations have application to a
user who wants to accurately compute his coordinate position and
time. In general, the world function approach is useful in
applications where high-accuracy measurements must be made over
large distances. An application of recent interest is the design
of space-based interferometers for precision sensing and
surveillance purposes~\cite{Thompson,GRACE}. In some of these
designs, in order to achieve high-resolution imaging long base
lines (hundreds of kilometers) must be used between Earth
satellites and their separations may need to be accurate to within
a micrometer or better. To unambiguously and accurately define
such positions, a curved space-time approach should be used that
takes into account the warping of the geometry of space-time due
to gravitational effects.

In section II, I write the generalization of Eq.\
(\ref{LightCone}) in terms of the world function and point out the
limitations of navigation in curved space-time by electromagnetic
beacons. In section III, I briefly describe the restricted problem
of computing an observer's coordinate time if his or her spatial
coordinates are known (a restricted type of navigation). In
sections IV and V, I give the detailed equations applicable to
navigation in the vicinity of the Earth in Schwarzschild
coordinates and in coordinates that rotate with the Earth.

\section{Navigation Equations}

The world function was initially introduced into tensor calculus
by Ruse~\cite{Ruse1931a,Ruse1931b}, Synge~\cite{Synge1931}, Yano
and Muto~\cite{YanoandMuto1936}, and Schouten~\cite{Schouten1954}.
It was further developed and extensively used by Synge in
applications to problems dealing with measurement theory in
general relativity~\cite{Synge1960}. In general, the world
function has received little attention, so I give the following
definition. Consider two points, $P_1$ and $P_2$, in a general
space-time, connected by a unique geodesic path $\Gamma$ given by
$x^i(u)$, where $u_1 \le u \le u_2$. A geodesic is defined by a
class of special parameters $u^\prime$ that are related to one
another by linear transformations $u^\prime = a u + b$, where $a$
and $b$ are constants.  Here, $u$ is a particular parameter from the
class of special parameters that define the geodesic $\Gamma$, and
$x^i(u)$ satisfy the geodesic equations
\begin{equation}
\frac{d^2 x^i}{du^2}+ \Gamma^i_{jk} \frac{dx^j}{du} \frac{dx^k}{du}
=0
\label{GeodesicDiffEq}
\end{equation}
The world function between $P_1$ and $P_2$ is defined as the
integral along $\Gamma$
\begin{equation}
\Omega(P_1,P_2) = \frac{1}{2} (u_2 - u_1) \int^{u_2}_{u_1} \, g_{ij} \frac{dx^i}{du}
\frac{dx^j}{du} \, du
\label{WorldFunctionDef}
\end{equation}
The value of the world function has a geometric meaning: it is one-half the
square of the space-time distance between points $P_1$ and $P_2$. Its value
depends only on the eight coordinates of the points $P_1$ and
$P_2$. The value of the world function in Eq.\
(\ref{WorldFunctionDef}) is independent of the particular special
parameter $u$ in the sense that under a transformation from one
special parameter $u$ to another, $u^\prime$, given by $u=a
u^\prime + b$, with $x^i(u)=x^i(u(u^\prime))$, the world function
definition in Eq.\ (\ref{WorldFunctionDef}) has the same form
(with $u$ replaced by $u^\prime$).

The world function is a two-point invariant in the sense
that it is invariant under independent transformation of
coordinates at $P_1$ and at $P_2$.  Consequently, the world
function characterizes the space-time. For a given space-time, the
world function between points $P_1$ and $P_2$ has the same value
independent of the metric-induced coordinates.  A simple example
of the world function is for Minkowski space-time, which is given
by
\begin{equation}
\Omega(x^i_1,x^j_2) = \frac{1}{2}\, \eta_{ij} \, \Delta x^i  \,  \Delta x^j
\label{MinkowskiWorldFunction}
\end{equation}
where $\eta_{ij}$ is the Minkowski metric with only non-zero
diagonal components $(-1,+1,+1,+1)$, and $ \Delta x^i  = (x_2^i -
x_1^i)$, $i=0,1,2,3$, where $x_1^i$ and $x_2^i$ are the
coordinates of points $P_1$ and $P_2$, respectively.  Up to a
sign, the world function gives one-half the square of the
geometric measure in space-time. Calculations of the world
function for specific space-times can be found in
Refs.~\cite{Synge1960,John1984,John1989,Buchdahl79} and
application to Fermi coordinates in Synge\cite{Synge1960} and
Gambi et al.~\cite{Gambi1991}.

The generalization to a curved space-time of the navigation Eq.\
(\ref{LightCone}) is given by
\begin{equation}
\Omega(x^i_s,x^j_o) = 0, \;\;\;\; s=1,2,3,4
\label{CovariantNavigation}
\end{equation}
where $x^i_s=(t_s,{\bf x}_s)$ are the coordinates of the emission events,
$x^i_o=(t_o,{\bf x}_o)$ are the observer coordinates and the world function is
defined by Eq.\ (\ref{WorldFunctionDef}). Within the geometric optics
approximation, Eq.~(\ref{CovariantNavigation}) forms a natural generalization of
the navigation Eq.\ (\ref{LightCone}).  In addition to Eq.\
(\ref{CovariantNavigation}), the appropriate causality conditions $t_s
> t_o$ or $t_o > t_s$, for $s=$1,$\dots$4, must be added.
The set of relations in Eq.\ (\ref{CovariantNavigation}) are manifestly covariant
and invariant due to the transformation properties of the world function under
independent space-time coordinate transformations at point $P_1$ and at $P_2$.

From the definition of the world function, the intrinsic
limitations of navigation in a curved space-time are evident: the
world function $\Omega(P_1,P_2)$ must be a single-valued function
of $P_1$ and $P_2$. In general, if two or more geodesics connect
the points $P_1$ and $P_2$, then $\Omega(P_1,P_2)$ will not be
single-valued and the set of equations in Eq.\
(\ref{CovariantNavigation}) may have multiple solutions or no
solutions.  Such conjugate points $P_1$ and $P_2$ are known to
occur in applications to planetary orbits and in
optics~\cite{Synge1960}. However, when the points $P_1$ and $P_2$
are close together in space and in time and the curvature of
space-time is small, we expect the world function to be single
valued and the solution of Eq.\ (\ref{CovariantNavigation}) to be
unique. Therefore, navigation in curved space-time is limited by
the possibility of determining a set of four unique null geodesics
connecting four emission events to one reception event. In the
case of strong gravitational fields as may exist in the vicinity
of a black hole, or when the (satellite) radio beacons are at
large distances from the observer in a space-time of small
curvature,  navigation by radio beacons may not be possible
in principle. In such cases, one may have to supplement radio
navigation by inertial techniques; see, for example, the discussion
by Sedov~\cite{Sedov1976}.

\section{Coordinate Time at a Known Spatial Position}

Consider the restricted problem of an observer that knows his spatial position and
wants to obtain his coordinate time~\cite{timeTransferComment}.  A null geodesic
connects the emission and reception events, so the value of the world function is
zero,
\begin{equation}
\Omega(t_s,{\bf x}_s,t_o,{\bf x}_o)=0
\label{timetransfer}
\end{equation}
where the satellite emission event coordinates $(t_s,{\bf x}_s)$
and the observer spatial coordinates ${\bf x}_o$ are known.
Equation~(\ref{timetransfer}) must be supplemented by the
causality condition $t_o> t_s$.  The observer obtains his coordinate time
by solving Eq.\ (\ref{timetransfer}) for $t_o$.

As a simple example of the application of Equation (\ref{timetransfer}),
consider an observer at a known spatial location who wants to compute his
coordinate time in flat space-time in a rotating system of coordinates, $y^i$, by
receiving signals from a satellite at $P_s=(t_s,{\bf y}_s)$.  I take the
transformation from Minkowski coordinates $x^i$ to rotating coordinates $y^i$
to be given by
\begin{eqnarray}
y^0 & = & x^0 \nonumber \\
y^1 & = & \cos(\frac{\omega}{c} x^0) \,
x^1 - \sin(\frac{\omega}{c} x^0) \, x^2 \nonumber \\
y^2 & = & \sin(\frac{\omega}{c} x^0) \, x^1 + \cos(\frac{\omega}{c} x^0) \, x^2 \nonumber \\
y^3 & = & x^3
\label{CoordinateTransformation}
\end{eqnarray}
The world function is a two-point invariant that characterizes the space-time,
so, in the rotating coordinates its value does not change.  Using the world function
for Minkowski space-time in Eq.\ (\ref{MinkowskiWorldFunction}) and the
transformation to rotating coordinates $y^i$ in Eq.\ (\ref{CoordinateTransformation}),
the world function is given by
\begin{eqnarray}\label{rotatingVacuumWorldFunction}
\Omega(P_s,P_o) & = & \Omega(t_s,{\bf x}_s,t_o,{\bf x}_o)  = \bar{\Omega}(t_s,{\bf
y}_s,t_o,{\bf y}_o) \nonumber \\
                                    & = & \frac{1}{2}\left[({\bf y}_o - {\bf y}_s)^2 - c^2
(t_o-t_s)^2\right] \nonumber \\
                                    &   & +   ({\bf y}_s \times {\bf y}_o)\cdot {\bf
                                    n}\,
\sin(\omega (t_o-t_s)) + 2 \left[ {\bf y}_s \cdot {\bf y}_o - ({\bf y}_s \cdot
{\bf n})({\bf y}_o \cdot {\bf n})  \right] \sin^2(\frac{1}{2}\omega (t_o-t_s))
\end{eqnarray}
where ${\bf n}$ is a unit vector in the direction of the angular velocity
vector ${\bf \omega}$, which I take to be along the z-axis. I assume that the angular
velocity of rotation is small, so that the time for light to travel from a satellite
at ${\bf y}_s$ to an observer at ${\bf y}_o$ is small compared to the period of
rotation $2\pi/\omega$.  I define the small dimensionless parameter $\delta=\omega
|{\bf y}_o - {\bf y}_s| /c << 1$.  Equation (\ref{timetransfer}) can then be
solved for $\Delta t = t_o - t_s$ by iteration, leading to
\begin{eqnarray}\label{y1y2DeltaTime}
c \Delta t & = & |{\bf y}_o - {\bf y}_s | + \frac{1}{c} ({\bf y}_s
\times {\bf y}_o)\cdot {\bf \omega} \nonumber \\
           &  & + \frac{1}{2c^2} |{\bf y}_o - {\bf y}_s | \left\{
           \frac{\left[ ({\bf y}_s \times {\bf y}_o)\cdot {\bf \omega} \right]^2}{|{\bf y}_o - {\bf y}_s|^2}+
\left[ \omega^2 {\bf y}_s \cdot {\bf y}_o - ({\bf y}_s \cdot {\bf \omega})({\bf y}_o \cdot {\bf
\omega}) \right] \right\}  +O(\delta^4)
\end{eqnarray}
The first term on the right side of Eq.\ (\ref{y1y2DeltaTime})
divided by $c$ is the time for light to travel from the emission
event at the satellite, $(t_s,{\bf y}_s)$, to the observer at
event, $(t_o,{\bf y}_o)$, in the absence of rotation.  The second
term is the celebrated Sagnac
effect~\cite{Sagnac,Post,AndersonEtAl,Tartaglia}, which depends on
the sense of rotation of the coordinates (sign of ${\bf \omega}$).
The third term is a higher order correction that is independent of
the sense of rotation; i.e., it is the same when ${\bf \omega}
\rightarrow -{\bf \omega}$.  This term is on the order of
5$\times$10$^{-14}$ s for satellite and Earth angular velocity of
rotation parameters appropriate to the GPS. Equation
(\ref{y1y2DeltaTime}) leads to the standard expression for the
Sagnac effect when we take the difference of propagation times for
clockwise ($\Delta T_-$) and counterclockwise ($\Delta T+$)
propagation of light along the limit of a sequence of tangents on
the perimeter of a circle, $\Delta T_+ - \Delta T_- = 4 {\bf
A}\cdot \omega/c^2$, where ${\bf A}$ is the included
area\cite{Sagnac,Post,AndersonEtAl,Tartaglia}. Note that the third
term in Eq.\ (\ref{y1y2DeltaTime}) does not contribute to the
difference of round-trip times, $\Delta T_+ - \Delta T_-$, so it
is not measurable in a Sagnac experiment. However, this term does
contribute to a determination of coordinate time.

\section{Navigation in the Vicinity of the Earth}

In the vicinity of the Earth, the gravitational potential can be
approximated by~\cite{Caputo1967}
\begin{equation}
\phi(r,\theta)= -\frac{G M}{r} \left[ 1 - J_2 \left( \frac{R_e}{r}
\right)^2 P_2(\cos(\theta))  \right] \label{EarthPotential}
\end{equation}
where $P_2(x) =(3 x^2 -1)/2$ is the second Legendre polynomial and
$J_2$ is the Earth's quadrupole moment, whose value is
approximately $J_2=1.0\times10^{-3}$.  However, for navigation in
the vicinity of the Earth, we can neglect $J_2$ since it is three
orders of magnitude smaller than the dimensionless coefficient of
the monopole potential $GM/r$, which already contributes small
corrections to propagation of electromagnetic radiation. I also
neglect the effects of the rotation of the Earth, which give rise
to small terms in the metric of space-time $g_{0\alpha}$, since
these effects are completely negligible at the present
time~\cite{Tartaglia}. Therefore, the
Earth's gravitational field can be sufficiently accurately
described using the Schwarzschild metric~\cite{LLClassicalFields}
\begin{equation}
-ds^2= -\left( 1 - \frac{2 G M}{c^2 r}\right)c^2 dt^2 +
\frac{dr^2}{1 - \frac{2 G M}{c^2 r}} + r^2(d\theta^2 +
\sin^2\theta d\phi^2 ) \label{SchwarzschildSolution}
\end{equation}
In Eq.\ (\ref{SchwarzschildSolution}), I neglect the gravitational field of the
sun and other planets, since the Earth is in free fall and these fields are
essentially (up to tidal terms) cancelled as a result of the equivalence
principle.

Using the transformation to rectangular coordinates
\begin{eqnarray}
x^0 & = & c t \nonumber \\
x^1 & = & r \sin \theta \cos \phi \nonumber \\
x^2 & = & r \sin \theta \sin \phi \nonumber \\
x^3 & = & r \cos \theta
\label{rectTransform}
\end{eqnarray}
and expanding in the small parameter $ GM/c^2 r$, the metric for
the Schwarzschild space-time can be written to first order as a sum of the
Minkowski metric, $\eta_{ij}$, and the deviation from
flatness tensor $ h_{ij}$ as
\begin{equation}
-ds^2 =  g_{ij} \, dx^i dx^j  =
\left( \eta_{ij} + h_{ij}\right) dx^i dx^j \label{Schwarzschild2}
\end{equation}
where $h_{ij} $ is given by
\begin{equation}
h_{ij} dx^i dx^j = \frac{2 G M}{c^2} \left[
\frac{(c dt)^2}{r} + \frac{(x^\alpha \, dx^\alpha)^2}{r^3} \right]
\label{htensor}
\end{equation}
The assumption that $GM/c^2 r \ll 1$ is a restriction on the
region of validity of Eq.\ (\ref{Schwarzschild2}) to large $r$
compared with the gravitational radius of the Earth, which is $2 G
M/c^2 \approx 0.88$cm.

Following Synge, I approximate the world function for the metric
in Eq.\ (\ref{Schwarzschild2}) by replacing the integrals over the
geodesic $\Gamma$ by integrals along a straight line, and taking
the special parameter $u$ to vary in the range $0\le u \le 1$,
which leads to~\cite{Synge1960}
\begin{equation}
\Omega(x^i_1,x^j_2) = \frac{1}{2}\, \eta_{ij} \, \Delta x^i  \,  \Delta x^j
+ \frac{1}{2}  \Delta x^i  \,  \Delta x^j \int_0^1 h_{ij} du
\label{SchwarzschildWorldFunction0}
\end{equation}
I find that explicit evaluation of these integrals (see Appendix B) leads
to~\cite{SyngeTypo}
\begin{eqnarray}
\Omega(x^i_1,x^j_2) & = & \frac{1}{2}\eta_{ij} \Delta x^i \Delta x^j +
\frac{GM}{c^2} \, \left[  |{\bf x}_2 - {\bf x}_1| +
   \frac{c^2 \Delta t^2}{|{\bf x}_2 - {\bf x}_1|} \right] \;
 \log \left( \frac{\tan(\frac{\theta_1}{2})}{\tan(\frac{\theta_2}{2})} \right)  \nonumber \\
  &  &  + \frac{GM}{c^2} |{\bf x}_2 - {\bf x}_1| \left(  \cos \theta_1 - \cos \theta_2
 \right)
\label{SchwarzschildWorldFunction}
\end{eqnarray}
where $c \Delta t = x^0_2 - x^0_1$, and  $\theta_1$ and $\theta_2$
are defined by
\begin{equation}
\cos \theta_a = \frac{{\bf x}_a \cdot ( {\bf x}_2 -  {\bf x}_1 ) }{|{\bf x}_a| |{\bf x}_2 - {\bf
x}_1|}, \;\;\;\; a=1,2
\label{cosineDef}
\end{equation}
See Appendix C for an estimate of the error in Eq.\
(\ref{SchwarzschildWorldFunction}). The first term on the right side of Eq.\
(\ref{SchwarzschildWorldFunction}) is the world function for Minkowski space-time,
given in Eq.\ (\ref{MinkowskiWorldFunction}). The second and third terms in Eq.\
(\ref{SchwarzschildWorldFunction}) give the corrections to the world function of
Minkowski space-time due to the gravitational effects of mass $M$. The expression
in Eq.\ (\ref{SchwarzschildWorldFunction}) can be used in Eq.\
(\ref{CovariantNavigation}) as a basis for navigation, or in Eq.\
(\ref{timetransfer}) for computing coordinate time in the vicinity of the Earth.

\begin{figure}[htbp]
\centerline{\epsfsize=6.0cm \epsfbox{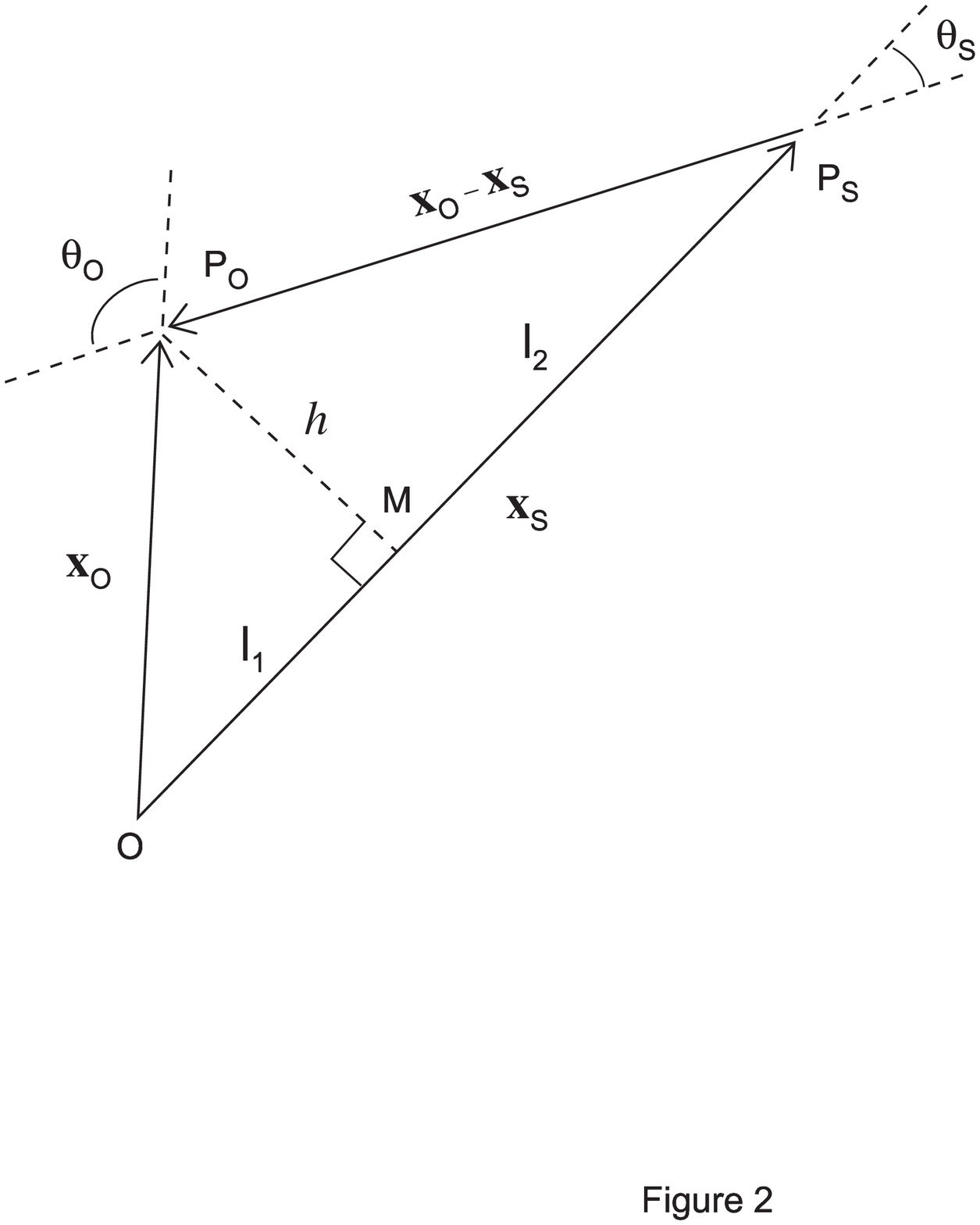}} \caption{The
spatial geometry of satellite at $P_S$ and receiver at $P_O$, with
coordinates ${\bf x}_s$ and ${\bf x}_o$, respectively, showing the
angles $\theta_s$ and $\theta_o$.}
\label{SatelliteReceiverGeometryFigure}
\end{figure}

As an example of using the world function in Eq.\
(\ref{SchwarzschildWorldFunction}), I consider determining the
coordinate time of an observer at a known spatial position ${\bf
x}_o$ in the vicinity of Earth, using standard Schwarzschild
coordinates. Taking the satellite position as ${\bf x}_s = {\bf
x}_1$,  the observer position ${\bf x}_o={\bf x}_2$, and making
use of the small parameter \mbox{$ GM/(c^2 |{\bf x}_o - {\bf
x}_s|) \ll 1 $}, I solve Eq.\ (\ref{timetransfer}) by iteration,
leading to
\begin{equation}
\label{timeTransferScwarzschild}
c \, \Delta t =  |{\bf x}_o - {\bf x}_s|
 + \frac{G M}{ c^2 }  \left[
 2 \log \left( \frac{\tan(\frac{\theta_s}{2})}{\tan(\frac{\theta_o}{2})} \right)
 +  \cos \theta_s - \cos \theta_o \right]
\end{equation}
The first term in Eq.~(\ref{timeTransferScwarzschild}) divided by
$c$ is the time for light to propagate from ${\bf x}_s$ to ${\bf
x}_o$. The second term is the small correction due to the presence
of the Earth's mass $M$ distorting the space-time in its vicinity.
This expression takes into account the delay of the
electromagnetic signal in a gravitational field (see, for example,
Ref.~\cite{Weinberg72,Will,WillGRReview1998Update} and references
cited therein). For a satellite directly overhead, where ${\bf
x}_o$ and ${\bf x}_s$ are co-linear with the origin of
coordinates, Eq.~(\ref{timeTransferScwarzschild}) leads to the
result
\begin{equation}\label{timeTransferScwarzschildOverheadSatellite}
c \, \Delta t =   |{\bf x}_o - {\bf x}_s|  + \frac{2 G M}{ c^2 }
 \log  \frac{|{\bf x}_s |}{|{\bf x}_o |}
\end{equation}
Equation Eq.~(\ref{timeTransferScwarzschildOverheadSatellite}) is obtained by
considering the spatial geometry shown in Figure
\ref{SatelliteReceiverGeometryFigure}. Define a triangle by the
three points: origin at $O$, satellite at $P_s$ with coordinates ${\bf
x}_s$, and observer at $P_o$ with coordinates ${\bf x}_o$.  From
point $P_o$, draw a line $\overline{P_o M}$ perpendicular to $\overline{OP}_s$
and define its length to be $h$.  Equation (\ref{timeTransferScwarzschildOverheadSatellite}) is
obtained from Eq.~(\ref{timeTransferScwarzschild}) by setting
$\theta_s=\theta_1$, $\theta_o=\theta_2$, ${\bf x}_s = {\bf x}_1$,
and ${\bf x}_o = {\bf x}_2$, using the definitions in
Eq.~(\ref{cosineDef})  and the relations
\begin{eqnarray} \label{cosEq}
\cos \theta_s & = & -1+\frac{1}{2} \, \left(\frac{h}{l_1}\right)^2 \\
\cos \theta_o & = & -1+ h^2 \, \left(\frac{1}{l_1 l_2} + \frac{1}{2 l_1^2}  + \frac{1}{2 l_2^2}\right)
\end{eqnarray}
where
\begin{eqnarray}  \label{lDEfs}
l_1 & = &  \left({\bf x}_s - {\bf x}_o \right) \cdot \frac{{\bf x}_s}{|{\bf x}_s|} \\
l_2 & = &  \frac{{\bf x}_o \cdot {\bf x}_s}{|{\bf x}_s|} \\
l_2^2 & = &  | {\bf x}_o |^{ \, 2} - h^2
\end{eqnarray}
and taking the limit as $h\rightarrow 0$.
Note that
Eq.~(\ref{timeTransferScwarzschild})
and~(\ref{timeTransferScwarzschildOverheadSatellite}) are
expressed in terms of Schwarzschild spatial coordinates (and do
not contain the temporal coordinates) of satellite ${\bf x}_s$ and
observer ${\bf x}_o$, because the Schwarzschild space-time is
static; i.e., the space-time admits a hypersurface orthogonal
time-like Killing vector field~\cite{DInverno}.

\section{Navigation in the Vicinity of the Earth in Rotating Coordinates}

In practical navigation problems, an observer or user of a
satellite navigation system is often interested in their
space-time position with respect to the Earth--which defines a
rotating coordinate system. To compute an observer space-time
position in a rotating system of coordinates, I apply the
navigation Eq.\ (\ref{CovariantNavigation}) in the rotating
system.  Having computed the world function for Schwarzschild
space-time in Eq.\ (\ref{SchwarzschildWorldFunction}) in a
`nonrotating' system of coordinates $x^i$, the invariant nature of
the world function can be used to write the world function in a
rotating system of coordinates, $y^i$, using the transformation in
Eq.\ (\ref{CoordinateTransformation}):
\begin{eqnarray}
\Omega(x^i_1,x^j_2)  & = &
\tilde{\Omega}(t_1,{\bf y}_1,t_2,{\bf y}_2) \nonumber \\
 & = & \frac{1}{2} \eta_{ij} \,
y^i y^j + \Delta F   \nonumber \\
   &    &  + \frac{G M}{c^2} \frac{1}{ \left[ ({\bf y}_2 -{\bf y}_1)^2 + 2 \Delta F
\right]^{1/2}} \left[  ({\bf y}_2 -{\bf y}_1)^2 + (c \, \Delta t)^2 + 2 \, \Delta F
\right] \, \log \left( \frac{F_2}{F_1} \right) \nonumber \\
  &   & - \frac{G M}{c^2} ( |{\bf y}_1|  + |{\bf y}_2|) +
\frac{G M}{c^2} \left( \frac{1}{|{\bf y}_1 |} +  \frac{1}{|{\bf y}_2 |} \right)
\left[ {\bf y}_1 \cdot {\bf y}_2  - \Delta F \right]
\label{rotatingSchwarzschildWorldFunction}
\end{eqnarray}
where
\begin{eqnarray}\label{dF}
\Delta F & = & ( {\bf y}_1 \times {\bf y}_2 )\cdot {\bf n} \, \sin(\omega \Delta t)
+ 2 \left[ {\bf y}_1  \cdot {\bf y}_2  -  ({\bf y}_1 \cdot {\bf n}) ({\bf y}_2 \cdot {\bf n}) \right]
\sin^2 \left(\frac{\omega \Delta t}{2} \right) \\
F_1 & = & {\bf y}_1 \cdot ({\bf y}_2 - {\bf y}_1 ) - \Delta F +
|{\bf y}_1 | \, \left[({\bf y}_2 - {\bf y}_1 )^2 + 2 \, \Delta F
\right]^{1/2} \\
F_2 & = & {\bf y}_2 \cdot ({\bf y}_2 - {\bf y}_1 ) + \Delta F +
|{\bf y}_2 | \, \left[({\bf y}_2 - {\bf y}_1 )^2 + 2 \, \Delta F
\right]^{1/2}
\end{eqnarray}

For navigation in the vicinity of the Earth in rotating
coordinates, the observer must solve the four simultaneous
Eqs.~(\ref{CovariantNavigation}) using the world
function in Eq.~(\ref{rotatingSchwarzschildWorldFunction}).  In
general, this must be done numerically.

Consider the simpler problem of determining an observer's
coordinate time at a known spatial location. Analytic results can
be obtained in this case. I solve Eq.~(\ref{timetransfer}) for
$\Delta t=t_o-t_s$ using the world function in
Eq.~(\ref{rotatingSchwarzschildWorldFunction}) by defining three
small dimensionless parameters, $x=\omega \Delta t$, $\delta=\omega
|{\bf y}_o - {\bf y}_s|/c$, and $\alpha=GM\omega/c^3$, and solving for
$x$ as a function of $\delta$ by iteration. This leads to
\begin{eqnarray}\label{dtRotatingSchwarzschild}
c \Delta t & = & |{\bf y}_o - {\bf y}_s| + \frac{1}{c} ( {\bf y}_s
\times {\bf y}_o )\cdot {\bf \omega} + \frac{1}{2 c^2} |{\bf y}_o
- {\bf y}_s| \left\{ \frac{\left[ ( {\bf y}_s \times {\bf y}_o
)\cdot {\bf \omega} \right]^2}{|{\bf y}_o -{\bf y}_s|^2} +
\omega^2 {\bf y}_s \cdot {\bf y}_o - ({\bf y}_s \cdot {\bf
\omega}) ({\bf y}_o \cdot {\bf \omega}) \right\} \nonumber  \\
  &  & + \frac{G M }{c^2} \left[  2 +
  \frac{1}{c} \frac{({\bf y}_s \times {\bf y}_o)\cdot {\bf \omega}}{|{\bf y}_o -{\bf
  y}_s|} \right]
  \log \left( \frac{\tan(\frac{\theta_o}{2})}{\tan(\frac{\theta_s}
  {2})} \right)
 + \frac{G M }{c^2} \left[  1 +
  \frac{1}{c} \frac{({\bf y}_s \times {\bf y}_o) \cdot {\bf \omega}}{|{\bf y}_o -{\bf
  y}_s|}  \right]
 \left(  \cos \theta_s - \cos \theta_o \right)
\end{eqnarray}
In Eq.~(\ref{dtRotatingSchwarzschild}), I have dropped small terms of $O(\alpha
\delta)$, $O(\alpha^2)$, and $O(\delta^4)$.  Equation
(\ref{dtRotatingSchwarzschild}) gives the coordinate time for the signal to travel
from the source to the observer, $\Delta t = t_o - t_s$.  Since the propagation
time $\Delta t$ is given in a rotating system of coordinates,
Eq.~(\ref{dtRotatingSchwarzschild}) contains the Sagnac effect, modified by the
presence of mass M (the Earth). The first term corresponds to the propagation of
the electromagnetic signal from the satellite at ${\bf y}_s$ to observer at ${\bf
y}_o$ in (flat, nonrotating) Minkowski space-time coordinates. The second term is
the standard Sagnac correction term (expressed in terms of coordinate time) that
depends on the sense of rotation ${\bf \omega}$, which appeared in rotating
coordinates in vacuum (see Eq.~(\ref{y1y2DeltaTime})). The third term on the right
side of Eq.~(\ref{dtRotatingSchwarzschild}) also appears in flat space-time
and, as previously mentioned, it does not depend on the sense of the rotation; it is
the same when $\omega \rightarrow -\omega$ and, hence, does not contribute to the
Sagnac effect. The last two terms on the right-hand side of
Eq.~(\ref{dtRotatingSchwarzschild}) are corrections to the coordinate time of
propagation due to the Earth's mass $M$. Note that the coefficients of each of
the last two terms also depend on $\omega$.  This represents a modification of the
Sagnac effect due to the presence of mass $M$.

The rotation of the coordinate system leads to a break in spherical three-dimensional
symmetry. Note that terms two and three on the right side of
Eq.~(\ref{dtRotatingSchwarzschild}),  which are due to the rotation, have a
cylindrical symmetry determined by the direction of ${\bf \omega}$, as expected.
On the other hand, the last two terms that depend on the mass $M$ have
coefficients that have a constant (spherically symmetric) term plus a term that
depends on the sense of rotation (linear in $\omega$).

\section{Conclusion}

If an observer simultaneously receives electromagnetic signals
from four electromagnetic beacons in flat space-time, he or she
can compute their position in space-time by solving the four light
cone Eqs.~(\ref{LightCone}). This procedure is routinely carried
out everyday by users of satellite navigation systems such as GPS
and GLONASS. Equation~(\ref{LightCone}) neglects small effects of
gravitational fields on electromagnetic signal propagation. In
this work, I have included these effects in a natural way using
the two-point invariant world function developed by J. L. Synge. I
have given a simple covariant and invariant formulation of the
navigation problem in Eq.\ (\ref{CovariantNavigation}).  An
approximation to the world function in Schwarzschild coordinates
is given in Eq.\ (\ref{SchwarzschildWorldFunction}), and in
coordinates that rotate with the Earth in Eq.\
(\ref{rotatingSchwarzschildWorldFunction}). In the future,
approximations to the world function may be obtained for the case
of the Eddington and Clark metric~\cite{EddingtonandClark1938} and
Eqs. (\ref{LightCone}) may serve as the basis for high-accuracy
navigation throughout our solar system.

I have implicitly used the geometric optics approximation by assuming that
electromagnetic radiation propagates along null geodesics. Furthermore, when
navigating inside the Earth's atmosphere, or when radiation traverses the
atmosphere, there is an additional signal delay that is bigger than the effects
discussed here and must be taken into account in detail.  I have neglected these
effects. Consequently, the results here are valid for observers in orbit around
the Earth outside the Earth's atmosphere.  Work is in progress to extend the
world function approach to simultaneously include gravitational and
index-of-refraction (atmospheric) effects.

The world function approach used here can be applied to areas that require precise
definitions of distances over satellite-to-ground length scales. A particular
application area of recent interest, where high-accuracy position is required over
hundreds of kilometers, is flying satellites in formation, for precise sensing and
surveillance applications. The high-accuracies needed in definition of
satellite positions may require a curved space-time theoretical framework that includes
gravitational corrections to flat Minkowski space-time.

\appendix
\section{Conventions and Notation}

Where not explicitly stated otherwise, I use the convention that Roman indices,
such as on space-time coordinates $x^i$ take the values $i=0,1,2,3$ and Greek
indices take values $\alpha=1,2,3$. Summation is implied over the range of the
index when the same index appears in a lower and upper position.  In some cases,
such as Eq.~(\ref{htensor}), Greek indices summation is implied when both indices
are in the upper position, such as in $x^\alpha dx^\alpha$.

If $x^i$ and $x^i + dx^i$ are two events along the world line of an ideal clock,
then the square of the proper time interval between these events  is $d\tau=ds/c$,
where the measure $ds$ is given in terms of the space-time metric as $ ds^2 =
-g_{ij}\, dx^i \, dx^j $.  I choose $g_{ij}$ to have the signature $+$2. When
$g_{ij}$ is diagonalized at any given space-time point, the elements can take the
form of the Minkowski metric given by $\eta_{00}=-1$, $\eta_{\alpha
\beta}=\delta_{\alpha \beta}$, and $\eta_{0 \alpha}=0$.

\section{Integrals}
Two integrals are needed to explicitly evaluate the world function in
Eq.~(\ref{SchwarzschildWorldFunction0}) (see Ref.~\cite{SyngeTypo}):
\begin{equation}
\int_0^1 \frac{1}{r}\, du = \frac{1}{|{\bf x}_1 -{\bf x}_2|}
\log \left( \frac{\tan(\frac{\theta_1}{2})}{\tan(\frac{\theta_2}{2})} \right)
\label{int1}
\end{equation}

\begin{equation}
\Delta x^\alpha \, \Delta x^\beta \int_0^1 \frac{x^\alpha \,
x^\beta}{r^3} \, du = |{\bf x}_1 -{\bf x}_2| \, \left[
\log \left( \frac{\tan(\frac{\theta_1}{2})}{\tan(\frac{\theta_2}{2})}
\right) + \cos(\theta_1) - \cos(\theta_2) \right]
\label{int2}
\end{equation}
where $\theta_1$ and $\theta_2$ are defined in Eq.\ (\ref{cosineDef}).

\section{Error in Approximation of the World Function}

In Eq.\ (\ref{SchwarzschildWorldFunction}), I use Synge's approximation to the
world function for the metric in Eq.\ (\ref{Schwarzschild2}), which entails
replacing the integrals over the geodesic $\Gamma$ by integrals along a straight
line L
\begin{equation}
\Omega(P_1,P_2) =   \frac{1}{2} (u_2 - u_1) \int_\Gamma \,
g_{ij} \frac{dx^i}{du} \frac{dx^j}{du} \, du \approx
 \frac{1}{2} (u_2 - u_1) \int_L \,
g_{ij} \frac{dx^i}{du} \frac{dx^j}{du} \, du
\label{WorldFunctionApproximation}
\end{equation}

I take the straight line L to be given by
\begin{equation}
x^i_L(u)= k (u_2-u) x^i_1+ k (u-u_1) x^i_2 \label{Line}
\end{equation}
where $u_1 \le u  \le u_2$ and  $k=(u_2-u_1)^{-1}$. The error made in this
approximation can be computed as follows. Consider two points, $P_1$ and $P_2$,
connected by a family of curves $x^i(u,v)$, where $u$ and $v$ are independent
parameters. I assume that $P_1=x^i(u_1,v)$ and $P_2=x^i(u_2,v)$, for all $v$. It
is convenient to define the two vector fields
\begin{equation}
U^i=\frac{\partial x^i}{\partial u}, \;\;\;\;\;\;\;\; V^i=\frac{\partial x^i}{\partial v}
\label{vectorfields}
\end{equation}
Note that by construction $V^i(u_1,v)=V^i(u_2,v)=0$. Furthermore, assume that a
unique geodesic $\Gamma$ connects the points  $P_1$ and $P_2$, and that this
geodesic $\Gamma$ is given by the curve with a particular value of $v=v_o$, namely
$x^i(u,v_o)$ for $u_1\le u \le u_2$.  Define the integral
\begin{equation}
I(v) =   \frac{1}{2} (u_2 - u_1) \int_{u_1}^{u_2} \, g_{ij} U^i U^j \, du
\label{I-integral}
\end{equation}
Now, assume that we have a space-time of small curvature and expand the integral
$I(v)$ about the geodesic $\Gamma$
\begin{equation}
I(v) = \Omega(P_1,P_2) + \frac{d I (v_o)}{d v} (v-v_o) +
\frac{1}{2} \frac{d^2 I (v_o)}{d v^2} (v-v_o)^2 + \cdots
\label{I-expansion}
\end{equation}
The second term, $d I (v_o)/d v$, vanishes since this is the definition of a
geodesic curve.  The error in replacing the integral over a geodesic by an
integral over a nearby curve can then be estimated by the third term on the right
side of Eq.\ (\ref{I-expansion}). Using the relations
\begin{equation}
\frac{\delta V^i}{\delta u} =\frac{\delta U^i}{\delta v}
\label{cross-derivatives1}
\end{equation}
and
\begin{equation}
\frac{\delta U^i \; (u,v_o)}{\delta u} =  0
\label{cross-derivatives2}
\end{equation}
I find the approximate error is given by
\begin{eqnarray}
I(v) - \Omega(P_1,P_2) & \approx & \frac{1}{2} \frac{d^2 I (v_o)}{d v^2} (v-v_o)^2
\nonumber \\
                       &   =     &
\frac{1}{2}(u_2-u_1)(v-v_o)^2 \; \int_{u_1}^{u_2} \left[ g_{ij} \frac{\delta
U^i}{\delta v} \frac{\delta U^j}{\delta v} + R_{abcd} V^a U^b U^c V^d
\right]_{v=v_o} \; d u \label{WorldFunctionApprox}
\end{eqnarray}
where the Riemann tensor is given by
\begin{equation}
R^i_{jkm} = \Gamma^i_{jm,k} -\Gamma^i_{jk,m} + \Gamma^a_{jm} \Gamma^i_{ak}
-\Gamma^a_{jk} \Gamma^i_{am}
\label{RiemannTensor}
\end{equation}
the affine connection is
\begin{equation}
 \Gamma^i_{jk} = \frac{1}{2} g^{il}\left( g_{jl,k} + g_{kl,j} - g_{jk,l} \right)
\label{connection}
\end{equation}
and ordinary partial derivatives with respect to the coordinates are indicated by commas.
In Eq.~(\ref{WorldFunctionApprox}), I have dropped terms of order $O(v-v_o)^3 $.

To estimate the error incurred in Eq.\
(\ref{SchwarzschildWorldFunction0}) by approximating the world
function integral over a geodesic by an integral over the straight line in Eq.\ (\ref{Line}),
I construct an explicit parametrization of curves connecting $P_1$ and
$P_2$:
\begin{equation}
x^i(u,v)=\frac{v_o-v}{v_o} x^i_L(u) + \frac{v}{v_o} x^i_{\Gamma} (u)
\label{Curve}
\end{equation}
where $x^i_L(u)$ is given by Eq.\ (\ref{Line}) and $x^i_{\Gamma} (u)$  is a
geodesic connecting the points $P_1$ and $P_2$. For simplicity, in  Eq.\
(\ref{Curve}) I am assuming that the parameter  $u=0$ at $P_1$ and  $u=u_2$ at
$P_2$.  For the geodesic connecting the points  $P_1$ and $P_2$, I use a series
solution of Eq.\ (\ref{GeodesicDiffEq})
\begin{equation}
x^i_{\Gamma}(u)= x_1^i + \left( \xi^i +  \frac{1}{2} \Gamma^i_{jk}(P_1) \xi^j \xi^k
\right)\frac{u}{u_2} -\frac{1}{2}  \Gamma^i_{jk}(P_1) \xi^j \xi^k
\left( \frac{u}{u_2} \right)^2 + \cdots
\label{SeriesSolnGeodesic}
\end{equation}
where $\xi^i = x^i_2 -x^i_1$.  Using the curve parametrization in  Eq.\ (\ref{Curve}),
and carrying out the required calculations,
I find an estimate of the error in approximating the world function integral by a straight line,
given by Eq.\ (\ref{WorldFunctionApprox}), to be
\begin{equation}
I(v) - \Omega(P_1,P_2) \approx \frac{1}{24}\left( \frac{v-v_o}{v_o}\right)^2
\eta_{ij} \,
 \Gamma^i_{ab}(P_1) \,  \Gamma^j_{cd}(P_1) \, \xi^a \xi^b \xi^c
\xi^d + O(3)
\label{WorldFunctionError}
\end{equation}
where third order terms have been dropped, and $ \Gamma^i_{ab}(P_1)$ is the
connection evaluated at $P_1$.  I have taken $\gamma_{ab,c}$ and $\gamma_{ab,cd}$
to be of $O(1)$.

\section{Note on Iterative Solution of Navigation Equations}

Equation~(\ref{CovariantNavigation}) is a set of four nonlinear algebraic
equations for the four coordinates $(t_o, {\bf x}_o)$.  A simple method of
solution can be applied by linearizing and solving the system by iteration.
Setting $x_o^j=x_o^j(n)+\delta x^j(n)$ in Eq.\ (\ref{CovariantNavigation}) and
expansion to first order in $ \delta \, x^j(n)$ gives a linear set of equations
for $\delta x^k(n)$
\begin{equation}
M^{(n)}_{a k} \, \delta x^k(n) = - \Omega(x^i_a,x^j_o(n)) = 0,
\;\;\;\; a=1,2,3,4 \label{MatrixEquations}
\end{equation}
where $ \delta x^k(n)$ is the correction to the n$^{th}$ trial value $x^j_o(n)$
and
\begin{equation}
M^{(n)}_{a k}  = \left[ \frac{\partial
\Omega(x^i_a,x^j_o)}{\partial x^j_o } \right]_{x^j_o=x^j_o(n)}
\label{Mmatrix}
\end{equation}

I start the iteration by making an ansatz $x^j_o(1)$ for user coordinates $x^j_o$,
and solve Eqs.\ (\ref{MatrixEquations}) for the first correction, $\delta x^k(1)$.
The improved solution is then taken to be $x^j_o(2) =x^j_o(1) + \delta x^k(1)$ and
substituted back into Eq.\ (\ref{MatrixEquations}).  Iteration is continued until
sufficiently small corrections $\delta x^k(n)$ are computed.



\begin{thebibliography}{99}
\bibitem{timeTransferComment}In the jargon of the timing and time disemination
community, determining time at a known location is referred to as ``time
transfer".
\end{thebibliography}
\end{document}